\documentclass[prl,twocolumn,aps]{revtex4}
\usepackage{verbatim}
\usepackage{graphicx}
\normalfont
\begin{document}


\title{Optically tunable nuclear magnetic resonance in a single quantum dot}


\author{M. N. Makhonin$^1$, E. A. Chekhovich$^{1,2}$, P. Senellart$^3$, A. Lema\^itre$^3$, M. S. Skolnick$^1$, A. I. Tartakovskii$^1$}

\address{$^{1}$ Department of Physics and Astronomy, University of Sheffield, S3 7RH,UK \\ $^{2}$ Institute of Solid State Physics, RAS, Chernogolovka, 142432, Russia,\\ $^3$ Laboratoire de Photonique et de Nanostructures, Route de Nozay, 91460 Marcoussis, France}
\date{\today}

\begin{abstract}

We report optically detected nuclear magnetic resonance (ODNMR) measurements on small ensembles of nuclear spins in single GaAs quantum dots. Using ODNMR we make direct measurements of the inhomogeneous Knight field from a photo-excited electron which acts on the nuclei in the dot. The resulting shifts of the NMR peak can be optically controlled by varying the electron occupancy and its spin orientation, and lead to strongly asymmetric lineshapes at high optical excitation. The all-optical control of the NMR lineshape will enable position-selective control of small groups of nuclear spins in a dot. Our calculations also show that the asymmetric NMR peak lineshapes can provide information on the volume of the electron wave-function, and may be used for measurements of non-uniform distributions of atoms in nano-structures.  
\\
\end{abstract}

\maketitle


Nuclear spins offer a nano-scale resource with extended spin life-times and coherence, leading to proposals to use nuclei for quantum computation \cite{Kane,Vandersypen,Leuenberger,Yusa} and coherent spin-memories \cite{Morton}. 
Strong interest in nuclear spin effects in semiconductors has also been recently  stimulated by research into manipulation of single spins in nano-structures, where the electron-nuclear (hyperfine) interaction plays an important role  \cite{Reilly,Xu,Brunner}. Direct control of nuclear spins by resonant techniques such as NMR is highly desirable for both electron and nuclear spin manipulation experiments. In the past NMR methods have been widely applied to large area  semiconductor structures (heterojunctions, quantum wells etc) containing very large number of nuclei in the range 10$^8$ or more \cite{Yusa,Barret,Eickhoff,Kempf}.  Further refinement of these methods has made it possible to detect magnetic resonance of as few as ~$10^4$ nuclei in otherwise abundant spin environments by detecting the optical response from individual GaAs quantum dot nano-structures \cite{Gammon1,Gammon2}. These micro-ODNMR experiments revealed strong dot-to-dot variation of resonant frequencies \cite{Brown1}, arising from interaction of small nuclear spin ensembles with random Knight fields from single spins of localized electrons \cite{Brown2}.  

In this work we take advantage of the strong gradients of the Knight field inside a quantum dot produced by the localized electron spin and enter a new regime of {\it nano}-ODNMR. By employing ODNMR techniques first reported in Refs.\cite{Gammon1,Gammon2}, we measure with high precision the Knight shifts in the resonant frequencies of each individual isotope spin sub-system in individual GaAs/AlGaAs interface dots and find their dependence on the polarization and power of optical excitation. By varying the optical power, we find striking modifications of the lineshape of the NMR spectrum of the dot. These arise from the Knight field variation across the dot which in turn is determined by the spatial distribution of the electron wave-function. The interpretations are supported by calculations, which further demonstrate that by employing the inhomogeneities of the Knight shifts, it becomes possible to access selectively, by appropriate resonant frequencies, small groups of nuclear spins located in different parts of the dot. This may be used for spatially-selective control of the nuclear spins in a nanometer-sized quantum dot. In addition, we show that the strong NMR frequency gradients induced by the localized electron may be used for detection with a few nm-resolution of non-uniform distributions of atoms inside nano-structures. 

The dependence of the NMR frequencies on the intensity and polarization of optical excitation arises from the optically-induced Knight field, $B_e$, a result of the contact hyperfine interaction between an individual nuclear spin and an electron confined in the dot \cite{Abragam,Lai,Chekhovich}. In an uncharged dot, as in our case, $B_e$ arises from the photo-excited electrons, with the time-averaged dot occupancy, $F$, and mean electron spin $s$, controlled by the intensity and polarization of light, respectively.  The time-averaged magnitude of the Knight field for a nucleus  with a hyperfine constant $A$ at the position $\mathbf{r}$ depends on the nuclear gyromagnetic ratio $\gamma$ and is given by \cite{Abragam}:
\begin{equation}
 B_e = {{v_0  A}\over{\hbar \gamma Z}} |{\psi(\mathbf{r})}|^2 s F \label{Be}
\end{equation}
Here $v_0$ is the volume of the crystal unit cell, containing $Z=4$ Ga or As nuclei. $B_e$ follows the distribution of   the electron envelope function $\psi(\mathbf{r})$ in the dot leading to the nuclear site-dependent field varying across the dot. In what follows the corresponding site-dependent Knight shifts in the NMR frequency given by $\Delta f=\gamma B_e/2\pi$ are measured with high precision in individual QDs.  

The sample investigated contains interface QDs formed naturally by 1 monolayer width fluctuations in a nominally 13 monolayer GaAs layer embedded in Al$_{0.33}$Ga$_{0.67}$As barriers (see growth details in Ref.\cite{Peter}).  In contrast to self-assembled quantum dot structures, the interface dots are formed by  lattice-matched GaAs and AlGaAs layers, leading to reduced strain and weak quadrupole effects, resulting in narrow NMR linewidths. This makes the interface dots an ideal test bed for future applications of ODNMR in III-V semiconductor nano-structures. Knight field effects detected by ODNMR are observable in a wide range of magnetic fields, in contrast to the all-optical detection confined to very low B-fields  \cite{Lai,Chekhovich}. 

The ODNMR setup is sketched in Fig.1a. External magnetic field $B_{ext}$ is applied in the Faraday geometry. Optical excitation is used for (i) pumping the nuclear spin via dynamic nuclear polarization (DNP)\cite{Gammon1,Gammon2,Lai,Chekhovich,Eble,Nikolaenko,Tartakovskii,Chekhovich1,Chekhovich} and (ii) to excite photoluminescence (PL) for measurements of the exciton Zeeman splittings in individual dots. The measurements were carried out at a temperature $T=$4.2 K. We use an excitation laser at 670 nm which generates electrons and holes in the quantum well (QW) states $\approx$ 130 meV above the QDs emission lines. PL was detected with a double spectrometer and a charge coupled device. As shown in Fig.1a, a coil was wound around the sample for RF excitation of the dots. The coil was excited by the output from a radio frequency (RF) generator and provided transverse magnetic fields $B_{RF}$ up to 0.6 Gauss.

\begin{figure}
\centering
\includegraphics[width=8cm]{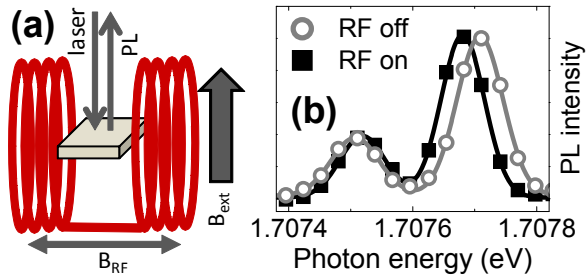}
\caption{(color online). (a) Diagram of the ODNMR experiment, depicting optical excitation and PL collection in the Faraday geometry, and the oscillating radio-frequency in-plane B-field $B_{RF}$. (b) PL spectra of the neutral exciton in a GaAs dot with and without RF excitation (black and gray symbols, respectively) at $B_{ext}=2$T. Lines show results of the peak fitting.}
\label{fig1}
\end{figure}

We study neutral dots. Fig.1b shows exciton PL spectra \cite{Nikolaenko} measured for $B_{ext} = 2$T under $\sigma^+$ polarized laser excitation with and without RF excitation (black and gray symbols, respectively). The two peaks observed in the spectrum belong to the exciton Zeeman doublet. Excitation with circularly polarized light results in the pumping of nuclear spins in the dot and gives rise to the Overhauser field $B_N$ \cite{Gammon1,Gammon2,Lai,Chekhovich,Eble,Nikolaenko,Tartakovskii,Chekhovich1}. $B_N$ is detected through the resulting change in the exciton Zeeman splitting, $\Delta E_{XZ}=g_e\mu_B B_N$ [$g_e$ electron g-factor, $\mu_B$ - Bohr magneton, $B_N$ is co-(anti-) parallel to $B_{ext}$ for $\sigma^{-}$($\sigma^{+}$) excitation]. Using lineshape fitting $\Delta E_{XZ}$ is  measured with an accuracy of $\approx$1$\mu$eV. RF excitation resonant with nuclear spin transitions in any of the three isotope sub­system contained in the dot ($^{75}$As, $^{71}$Ga, $^{69}$Ga \cite{Gammon1,Gammon2}) leads to nuclear spin depolarization and consequent reduction of $|B_N|$. This is observed in Fig.1b as a change in the splitting of the Zeeman doublet when RF excitation is applied. In what follows we will use the variation of the exciton Zeeman splitting upon optical or RF excitation to measure changes  in the nuclear polarization on the dot. 

Fig.2 shows NMR spectra $E_{XZ}(f_{RF})$ for $^{75}$As, $^{69}$Ga, $^{71}$Ga at $B_{ext} = 2$T recorded under simultaneous RF and circularly polarized laser excitation (a moderate pumping power, $P$ of 0.5 $\mu$W is employed, corresponding to $F\approx 0.1$). The spectra exhibit peaks and dips for $\sigma^-$ and $\sigma^+$ excitation, respectively, with a typical width of 8-10 kHz. Fig.2 shows a strong dependence of the resonance frequency on the polarization of optical excitation  \cite{Brown2}. A change of the laser polarization leads to the change in the time-averaged electron spin $s$ in Eq.\ref{Be}. When the polarization is tuned from $\sigma^+$ to $\sigma^-$, the resonance frequency is also tuned gradually as shown in Fig.2d. The total frequency shift between the resonances measured for $\sigma^+$ and $\sigma^-$ excitation corresponds to twice the maximum average Knight field $2B^{max}_e$ for a given optical power for nuclei of a particular isotope. For the moderate excitation power used in this experiment the following magnitudes of $B^{max}_e$ were found: $B_e(^{75}As) = 1$mT, $B_e(^{69}Ga) = 0.57$mT and $B_e(^{71}Ga) = 0.62$mT.  The above magnitudes of the Knight field are comparable with those reported for InGaAs QDs \cite{Lai} ($<$1mT) and somewhat smaller than up to 3 mT observed in InP dots \cite{Chekhovich}. Each individual field magnitude reported in Fig.2 is at least an order of magnitude larger than was reported for GaAs/AlGaAs QWs \cite{Eickhoff}, due to the stronger localization of the electron in quantum dots.  

\begin{figure}
\centering
\includegraphics[width=9cm]{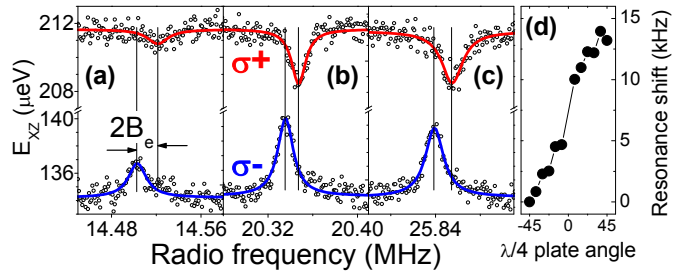}
\caption{(color online). Symbols show NMR spectra $E_{XZ}(f_{RF})$ measured at moderate optical power of 0.5 $\mu$W at $B_{ext}=2$T for $^{75}$As (a), $^{69}$Ga (b), $^{71}$Ga (c). Data obtained with $\sigma^{+(-)}$ excitation are fitted with the red (blue) line. Vertical lines mark positions of the NMR peaks. (d) Shift of the resonance frequency for $^{69}$Ga as a function of the rotation angle of the $\lambda/4$ plate in the laser excitation path.}
\label{fig1}
\end{figure}

We will now demonstrate the effect of electron occupancy $F$ (see Eq.\ref{Be}), increasing with the optical power, on the nuclear magnetic resonance frequency. Fig.3 shows NMR spectra recorded at $B_{ext}=2.5$ T using continuous RF and optical excitation. The spectra presented are measured for $^{69}$Ga with $\sigma^+$ polarized light. The vertical line and arrow at $f_0=25.812$MHz show the peak position of an NMR spectrum measured in the ''dark'' using pulsed techniques, where the laser is switched off during the RF excitation leading to $B_e=0$ \cite{pulsed}. A peak shift from the position depicted by the line  corresponding to $B_e=0$ is observed when the spectra are measured in the continuous excitation mode. This shift increases from 1 kHz for 0.1 $\mu$W excitation power to 7 kHz for 3 $\mu$W. In addition to the shift there is a strong modification of the lineshape: the full width at half maximum changes from ~8 kHz at 0.1 $\mu$W to 18 kHz at 3 $\mu$W. A pronounced asymmetry of the NMR spectrum is also observed at 3 $\mu$W, with the intensity decreasing to zero at a shift as high as 30 kHz.

In order to analyze the effect of the inhomogeneous Knight field quantitatively we have calculated the distribution of the NMR frequencies in an ensemble of nuclei in a dot under optical excitation. The position-dependent frequency shifts are calculated using Eq.\ref{Be} and can be expressed as $\Delta f(\mathbf{r}) = {v_0\over hZ}|{\psi(\mathbf{r})}|^2 A_i s F$. In GaAs $A_{As}=5.69$GHz, $A_{^{69}Ga}=4.66$GHz and $A_{^{71}Ga}=6.00$GHz.  For the spin ensemble of $^{69}$Ga (data in Fig.3) we use an experimentally observed Gaussian distribution of frequencies for an empty dot $F=0$ [i. e. $\Delta f(\mathbf{r})=0$] with a central frequency of $f_0=25.812$ MHz and a finite width $w_0=8$kHz most probably originating from weak residual strain. Under optical excitation, at a given radio frequency the response of the nuclei in the dot will be described with a Gaussian distribution where $f_0$ is replaced by the position-dependent $f_0+\Delta f(\mathbf{r})$.   

\begin{figure}
\centering
\includegraphics[width=8cm]{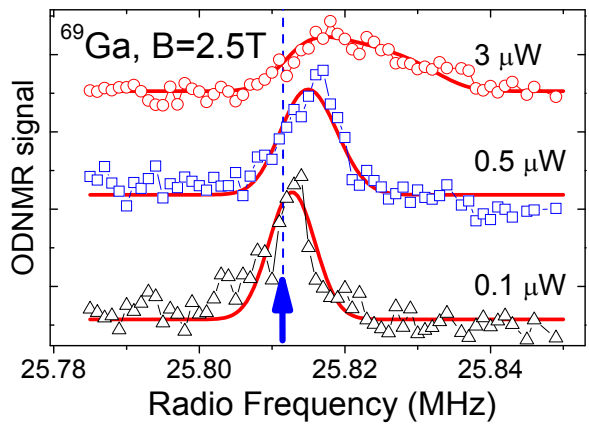}
\caption{(color online).  Symbols show NMR spectra $\Delta E_{XZ}(f_{RF})$ for $^{69}$Ga measured at $B_{ext}=2.5$T for powers 0.1, 0.5 and 3 $\mu$W. Lines show fitting obtained using Eq.\ref{shape} for a dot with the height of 4 nm and diameter 50 nm and the fitting parameter $sF$ of 0.02, 0.09 and 0.24, respectively. Arrow and dotted vertical line indicate position of the NMR peak measured in the dark \cite{pulsed}.}
\label{fig2}
\end{figure}

The Overhauser shift, $\Delta E_{XZ}$, observed in the experiment is a result of contributions from all nuclei in the dot. However, the contribution of individual nuclei  is dependent on the strength of the hyperfine interaction with the optically excited electron, and is therefore proportional to  $|{\psi(\mathbf{r})}|^2$. In the ODNMR experiment we measure the reduction in $|\Delta E_{XZ}|$ as a result of the interaction with RF excitation at a given frequency. Such depolarization of the nucleus at the position $\mathbf{r}$  will be observed at the resonant condition $f=f_0+\Delta f(\mathbf{r})$. Thus $|\Delta E_{XZ}(f)|$ can be calculated by integration over the volume of the dot:
\begin{equation}
 |\Delta E_{XZ}(f)| \propto \int{ |\psi(\mathbf{r})|^2 
 \exp[-(f-f_0-\Delta f)^2/2w^2_0] d^3r} \label{shape}
\end{equation}

The results of the fitting obtained with Eq.\ref{shape} are shown in Fig.3 by the full lines. For this model calculation we use an electron wave-function calculated for a cylindrical GaAs/Al$_{0.3}$Ga$_{0.7}$As dot of 4 nm height and 50 nm diameter with a uniform distribution of Ga and As. In the fitting procedure we also assume that $F\propto I_{PL}$, the PL intensity measured in experiment. As $I_{PL}$ saturates at high power, the magnitude of $sF$ obtained from the fitting in Fig.3 increases sub-linearly with the power: it is 0.02 for 0.1$\mu$W, 0.09 for 0.5$\mu$W, and 0.24 for 3$\mu$W. The maximum $sF$ of 0.24, obtained from the fitting is consistent with for example  $F\approx 0.5$ and $s\approx 0.5$, close to the experimental values. Note that alternatively we can use analysis of the data in Fig.3 to estimate the characteristic volume occupied by the electron wave-function, since both $s$ and $F$ can be estimated with good accuracy from the PL data.

The strong electron confinement leads to  strong gradients of the Knight field in the dot. To illustrate this point, Fig.4 shows the calculated Knight shift as a function of the in-plane distance $r$ (Fig.4a) and vertical coordinate $z$ (Fig.4b) from the position of the maximum of the electron wave-function. Here $sF=0.24$ and the wave-function is as in the fitting described above. The vertical bar in Fig.4b shows for comparison the width $w_0$ of the $^{69}$Ga resonance without the influence of the optically excited electron (the case of $F=0$). As seen from the plots, the presence  of the electron perturbs the resonance frequency distribution in a major way, leading to strong gradients in both in-plane and vertical directions: the resonance frequency changes by more than the $w_0$ on a 10 nm length-scale in-plane and in 2 nm in the  $z$-direction. This opens up the possibility for high spatial selectivity in addressing small groups of nuclei by choosing the appropriate frequency of resonant cw and pulsed RF excitation. To illustrate this possibility, Fig.4c shows a calculated color-plot representing an NMR-frequency map of the dot (for $sF=0.24$ and the wave-function used for the fitting in Fig.3), with individual colors showing locations of  nuclei with the same resonant frequency.  

\begin{figure}
\centering
\includegraphics[width=9cm]{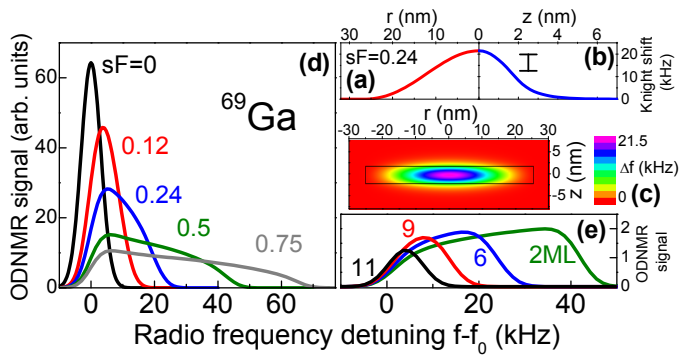}
\caption{(color online). The Knight shift as a function of the in-plane distance $r$ (a) and vertical coordinate $z$ (b) from the maximum of the electron wave-function for the case $sF=0.24$. (c) The distribution of the Knight shift in the nuclear spin ensemble in the dot (see the scale for $\Delta f$ opposite). The graph depicts a vertical  cross-section through the middle of the dot, with the dot boundaries shown by black lines. (d) NMR curves calculated for $^{69}Ga$ in the dot using Eq.\ref{shape} for different magnitudes of $sF$. (e) NMR curves calculated for $sF=0.5$ for a mono-layer of $^{69}Ga$ displaced vertically from the center of the dot by 2, 6, 9 and 11 mono-layers.}
\label{fig3}
\end{figure}

The NMR frequency gradients, and hence, spatial selectivity in manipulation of nuclear spins can be enhanced even further in the case with a higher magnitude of $sF$, as shown in Fig.4d: for $sF=0.75$ the width of the resonance can be increased by a factor exceeding 6 compared with an electron-free dot ($sF=0$). These can be realized in an electron-doped dot in high magnetic fields at low temperatures, where $F=1$ and $s$ may also approach 1. 

In the case when the electron wave-function can be measured, e. g. using magneto-transport measurements \cite{Vdovin}, it should also be possible to relate the given RF frequency to a small group of nuclei distributed on a contour with known spatial coordinates corresponding to equal electron probability. According to Fig.4a and b showing the Knight field gradients, in a dot the resolution of such a nano-ODNMR imaging method, provided solely by the inhomogeneity of the electron Knight field, will be better than 10 nm. Additional spatial direction-specific information (i.e. a form of magnetic resonance imaging) can be obtained by applying in-plane and vertical electric fields, leading to predictable modifications of the electron wave-function. 

Such non-invasive  microscopy method can find applications in structural studies of nano-structures containing non-uniform distributions of different types of atoms. 
To demonstrate the principle, we show in Fig.4e calculated NMR curves for a mono-layer (ML) of $^{69}$Ga situated 2, 6, 9 and 11 MLs (1 ML$=0.28$ nm) from the  the dot center in the case of an $sF$ factor of 0.5. A striking dependence of the resonance shapes on the ML position is observed with pronounced broadening and asymmetric shapes for the 2 and 6 ML displacement, where the peak maxima are observed at high Knight shifts. This is in contrast to the NMR lineshape for $sF=0.5$  in Fig.4d, indicating a smaller relative contribution of nuclei in the regions with low $|\psi(\mathbf{r})|^2$ in a ML compared to the whole dot.

To conclude, we demonstrate optically controlled tuning of the NMR frequencies of small ensembles of nuclear spins inside a semiconductor quantum dot. We have employed strain-free GaAs/AlGaAs interface QDs, where inhomogeneities of the resonant frequencies due to the non-zero nuclear quandrupole moments are not significant, in contrast to self-assembled QDs \cite{Maletinsky}.  We have been able to demonstrate tunability of the magnetic resonance in a dot by introducing resonant frequency distributions in a controlled fashion via optical excitation of a spin-polarized electron. This has potential for precise manipulation of Overhauser fields on the nano-scale, coherent manipulation of groups of only a few hundred of nuclear spins in GaAs nano-structures, and magnetic resonance imaging with a few nm resolution.

We thank K. V. Kavokin and A. J. Ramsay for fruitful discussions. This work has been supported by  the EPSRC Programme Grant EP/G601642/1. AIT is grateful for support by an EPSRC ARF.


\begin{thebibliography}{}

\bibitem{Kane} B. E. Kane, Nature {\bf 393}, 133–137 (1998)
\bibitem{Vandersypen} L. M. K. Vandersypen {\it et al.},  Nature {\bf 414}, 883–887 (2001).
\bibitem{Leuenberger} M. N. Leuenberger {\it et al.}, Phys. Rev. Lett. {\bf 89}, 207601 (2002).
\bibitem{Yusa} G. Yusa {\it et al.}, Nature {\bf 434} 1001 (2005).
\bibitem{Morton} J. J. L. Morton {\it et al.}, Nature {\bf 455} 1085 (2008).
\bibitem{Reilly} D. J. Reilly {\it et al.}, Science {\bf 321}, 817 (2008).
\bibitem{Xu} X. Xu {\it et al.}, Nature {\bf 459}, 1105 (2009).
\bibitem{Brunner} D. Brunner {\it et al.}, Science {\bf 325}, 70 (2009).
\bibitem{Barret} S. E. Barret {\it et al.}, Phys. Rev. Lett. {\bf 72}, 1368 (1994).
\bibitem{Eickhoff} M. Eickhoff {\it et al.}, Phys. Rev.{\bf B 71}, 195332 (2005).
\bibitem{Kempf} J. G. Kempf, M. A. Miller, D. P. Weitekamp, Proceedings of the National Academy of Sciences, {\bf 105}, 20124 (2008).
\bibitem{Gammon1} D. Gammon {\it et al.}, Science {\bf 277}, 85 (1997).
\bibitem{Gammon2} D. Gammon {\it et al.}, Phys. Rev. Lett. {\bf 86}, 5176 (2001).
\bibitem{Brown1} S.W. Brown {\it et al.}, Solid State Nuclear Magnetic Resonance {\bf 11}, 49 (1998). 
\bibitem{Brown2} Knight shifts in interface dots have been reported in Ref.\cite{Brown1}. However, no detailed investigation of their dependence on the power and polarization of optical excitation has been carried out. 
\bibitem{Abragam} A. Abragam, {\it Principles of Nuclear Magnetism}, Oxford University
Press (1961).
\bibitem{Lai} C. W. Lai {\it et al.}, Phys. Rev. Lett. {\bf 96}, 167403 (2006).
\bibitem{Chekhovich} E. A. Chekhovich {\it et al.}, arXiv:0901.4249, (2008).
\bibitem{Peter} E. Peter {\it et al}, Phys. Rev. Lett. {\bf 95}, 067401 (2005).
\bibitem{Eble} B. Eble {\it et al.}, Phys. Rev. B {\bf 74}, 081306(R) (2006).
\bibitem{Nikolaenko} A. E. Nikolaenko {\it et al.}, Phys. Rev. {\bf B 79}, 081303(R) (2009).
\bibitem{Tartakovskii} A. I. Tartakovskii {\it et al.}, Phys. Rev. Lett. {\bf 98}, 026806 (2007).
\bibitem{Chekhovich1} E. A. Chekhovich {\it et al.}, Phys. Rev. Lett. {\bf 104}, 066804 (2010).
\bibitem{pulsed} The sample is pumped with a train of 10 s pulses of circularly polarized light interrupted by 500 ms ''dark'' gaps. When light is switched off the dot is excited with a 500 ms RF pulse. PL is measured in the first 50 ms of each optical pulse. NMR peaks are obtained by varying the radio frequency and repeating the measurement cycle. 
\bibitem{Vdovin} E. E. Vdovin {\it et al.}, Science {\bf 290}, 122 (2000).
\bibitem{Maletinsky} P. Maletinsky {\it et al.}, Nature Physics {\bf 5}, 407 (2009).
 
\end{thebibliography}
\end{document}